# Annotating anatomy and pathology in the National Lung Screening Trial computed tomography images

Deepa Krishnaswamy, Vamsi Thiriveedhi, Suraj Pai, David Clunie, Igor Octaviano, Christopher P. Bridge, Steve Pieper, Ron Kikinis, Andrey Fedorov

*Corresponding author: Deepa Krishnaswamy (dkrishnaswamy@bwh.harvard.edu)*

# Abstract

Large-scale public medical imaging datasets contribute critically to translational research. When accompanied by rich clinical and multi-omics data, they can stimulate exploratory research and enable secondary analyses. Expert annotations of such imaging collections can support the development of new image analysis tools. Continuous enrichment of images with image-derived data makes them more usable for researchers without expertise in image analysis or access to large-scale computational resources. The National Lung Screening Trial (NLST) released a rich longitudinal dataset that includes Computed Tomography (CT) images for over 26,000 patients. We introduce three Digital Imaging and Communications in Medicine (DICOM) formatted datasets, complementing NLST CT images, shared as analysis results in the National Cancer Institute Imaging Data Commons (IDC). Two of those (IDC NLSTSeg and IDC NLST-Sybil) contain DICOM-harmonized annotations and extracted measurements (for 581 and 601 NLST patients, respectively) shared earlier using research formats (Sybil and NLSTseg). The third one (TotalSegmentator-CT-Segmentations) contains volumetric segmentations generated using TotalSegmentator and radiomics features for each segment for 26,194 NLST patients.

# Background & Summary

Publicly available cancer imaging datasets contribute critically to clinical research and the associated development and refinement of new analysis tools. Such datasets can help researchers to formulate and explore new hypotheses, facilitate opportunistic screening, and correlate image features with various clinical outcomes. In the field of artificial intelligence (AI), publicly available data enables investigators to develop and validate new methods, perform benchmarking against other models, and improve reproducibility. Unlike private or institutional datasets available through the approval process, public imaging datasets enable researchers to collaborate indirectly by continuously enriching the images with the analysis results, annotations, and image-derived measurements, which in turn can prompt new studies and simplify data reuse for new purposes, including by researchers without imaging expertise.

One of the largest publicly available datasets for lung cancer computed tomography (CT) data was produced by the National Lung Screening Trial (NLST), which was designed to study the utility of CT as the screening modality for lung cancer. The CT images collected in the course of this trial for over 26,000 patients from a wide distribution of scanners and institutions are publicly available, accompanied by rich clinical data and digital pathology slides for a subset of cancer-positive patients[1,2]. Since its release, the NLST collection has been utilized in numerous studies ranging from exploring new hypotheses for the NLST population[3,4] to developing or validating AI tools applicable beyond this cohort[5–7]. While clinical data accompanying the NLST collection includes information about the general location of the tumor in the cancer-positive subset, precise annotation of the tumor location or its margins was not available as part of the dataset. Studies often necessitate additional annotation of the images or lead to the creation of new analysis artifacts; those image-derived items are rarely shared.

However, three recent publicly released NLST annotation datasets challenged the *status quo*. Mikhael et al. released the Sybil[8] dataset in 2023 with the bounding box annotations of the NLST lesions, along with an AI model trained to predict an individual's risk of developing lung cancer. Independently, Chen et al. published the NLSTseg[9] in 2025, containing voxel-level volumetric segmentations of the lesions prepared by domain experts. While both of those datasets have been available for a year or more, they were not well-integrated with the imaging component of the NLST collection. Lastly, Bodard et al. released expert lesion annotations of NLST, where 20 radiologists independently annotated the data, and ensured that a case was read by at least four experts[10,11]. In total, the authors provided 2,004 DICOM SEG objects for 501 patients, and an annotations CSV file holding lesion characteristics.

The NLST CT images are available in the DICOM format, the original representation of the data as generated by the imaging equipment during the clinical trial, but de-identified to safeguard patient privacy[12]. Examining some of the typical use cases highlights limitations of the formats used for the original Sybil and NLSTseg datasets. *Visualization* of annotations is perhaps one of the first steps to assess their quality. There are no off-the-shelf tools to visualize Sybil annotations stored in a custom JSON representation. NIfTI format used by NLSTseg is broadly supported in research visualization tools, but it does not offer standard provisions for linking individual segmentations to the corresponding images. Neither of the formats is supported by off-the-shelf image hosting services that offer a standardized interface for accessing them. At the same time, numerous browser-based and desktop applications (both open-source and commercial) support images, and increasingly, segmentations and annotations encoded in DICOM[13]. The ability to *archive* images alongside image-derived content is another important use case, particularly crucial while managing large datasets that combine expert and AI-generated annotations. There are both commercial and open-source implementations of DICOM servers that can be used to organize DICOM-encoded data, enabling efficient and standardized search and retrieval via the DICOMweb interface[14]. Image-derived data encoded using non-standard conventions is not interoperable with such archives. Rich metadata, standardized semantics, and unique identifiers are critical to enable *search, cohort building, and linking* of the data. DICOM defines a unified data model that covers both images and image-derived content, object-specific metadata attributes and value sets, relies on SNOMED[15] for describing anatomy and findings, and documents conventions for assigning unique identifiers and utilizing those for establishing provenance of image-derived data. Instead, non-standard representations rely on custom conventions, e.g., dataset-specific JSON attributes in Sybil, or file names and documentation linking NIfTI files with the segmentation metadata provided in a separate spreadsheet. These non-standard encodings do not allow for establishing consistent conventions for metadata describing image-derived content, and can be error-prone. While such dataset-specific conventions can be manageable to address using custom tools while working with a single dataset, they lead to challenges aggregating data from multiple sources, redundant work to develop dataset-specific adapters, and a sacrifice of interoperability. Finally, DICOM is the required format for depositing data into the National Cancer Institute Imaging Data Commons (IDC),[16] the designated component of the US cancer data ecosystem for sharing images and image-derived data.

In this work, we present three image-derived DICOM datasets that complement NLST CT images: 1) the TotalSegmentator[17] anatomy segmentations and segmentation-derived radiomics features, 2) bounding box tumor annotations from Sybil[8], and 3) pixel-level segmentations of lesions from NLSTseg[9]. We demonstrate how the use of the DICOM standard implements the Findable, Accessible, Interoperable, and Reuse (FAIR) principles for data management[18]. We utilize the Segmentation (SEG) object to encode the pixel-level annotations, and Structured Reports (SRs) to encode both the slice-level bounding box annotations and segmentation-derived features. Our dataset is accompanied by Python notebooks to demonstrate the use of the resulting data, along with interactive dashboards that highlight the advantages of the interoperable representation of the released datasets.

# Methods

We curate and generate new data that augments the CT images available in the NLST collection. This data includes 1) secondary datasets that harmonize earlier released expert annotations of lung lesions (Sybil[8] and NLSTseg[9]) into a standard representation; 2) AI-generated annotations of the anatomic organs; 3) quantitative features extracted from the volumetric segmentations.

# NLST CT collection

The NLST collection contains data for over 26,000 patients enrolled in a clinical trial investigating the utility of low-dose CT for lung cancer screening (NCT 00047385, patient enrollment 2002-2004)[1,19]. NLST is one of the largest publicly available imaging datasets. With up to three time points for each patient, it contains over 200,000 CT scans accompanied by rich clinical data and digital pathology slides for a subset of patients. Radiology, digital pathology images, and a subset of clinical data (including patient characteristics, lung cancer staging, and pathology information) are available publicly[2] from The Cancer Imaging Archive (TCIA)[20] and Imaging Data Commons (IDC)[16], while complete clinical data can be requested from the NCI Cancer Data Access System (CDAS). NLST CT images were de-identified by TCIA[12] and are available in the standard DICOM format, which is the original representation generated by the CT scanner.

# DICOM harmonized image-derived data for NLST CT images

## Standardized DICOM objects

The DICOM standard[13] defines the conventions to encode, store, view, and transmit imaging and imaging-derived data. Within this standard, a number of objects, including DICOM SEG and SRs, cover the representation of image-derived data. DICOM SEG is a standard DICOM object that can be used to store voxel-level segmentations and allows for encoding metadata describing the segmentation (i.e., the semantics of the segmented region, details about how the segmentation was created, and references to the image being segmented). DICOM SRs, on the other hand, are used to encode readable text, graphical primitives annotating the image (e.g., points or bounding boxes), measurements derived from various annotations (both volumetric and planar), or clinical observations about the image findings, organized into a hierarchical tree of content. SRs rely heavily on standard ~~coded~~ terms to improve the semantic interoperability of the content and can reference related evidence, such as images or image-derived objects, establishing the machine-readable provenance chain. DICOM SR templates are defined in Part 16 of the standard[21], establishing patterns of this encoding tailored to a specific use case[22] (e.g., along with the general-purpose templates, DICOM defines domain-specific templates for such areas as ophthalmology, prostate MRI reporting, and mammography). The DICOM SR Template 1500 “Measurement Report” (TID1500)[23], is one of the generic templates that defines the standard DICOM representation for encoding arbitrary qualitative or quantitative measurements, which might be derived from planar annotations or volumetric segmentations[24–26]. The use of standard DICOM SEG and SR objects helps achieve FAIR[18] representation of the image-derived data, providing concrete standard means to encode metadata describing the content, including unique identifiers to support indexing and referencing of the related content, and maintain a composite context of the annotation (i.e., includes patient- and DICOM study-level attributes, thus linking it in the context of the other relevant DICOM data).

To implement the conversion and demonstrate the use of the dataset, we used the following open-source tools:

- *highdicom*[27] (version 0.27.0), a Python package that provides high-level abstractions for working with DICOM objects
- *dcmqi*[28] (version 1.4.0), a C++ library for conversion between imaging research formats and the standard DICOM representation for image analysis results
- *dcm2niix*[29,30], a command-line converter from DICOM to NIfTI format, which we used for DICOM CT data.

## NLST CT Annotations

### Sybil

The Sybil dataset[8] contains manual annotations of suspicious lung lesions in the CT scans corresponding to NLST patients who screened positive for cancer within a year after the CT scan. Annotations from the original dataset were prepared jointly by two fellowship-trained thoracic radiologists[8] using the MD.ai commercial software and were shared as per-slice bounding boxes stored in a single JSON file. We used the annotations available in the Sybil software release 1.6.0[31]. The original dataset was saved as a single JSON file with the tumor bounding boxes for all of the annotated images (where the bounding box is defined by its top-left corner and dimensions). Correspondence with the annotated CT images was captured in the JSON representation using DICOM SeriesInstanceUID and SOPInstanceUID identifiers, which uniquely identify the NLST CT image slice within that series.

To harmonize the representation of the Sybil annotations, we use DICOM SR TID1500, which enables lossless encoding of the original content. TID1500 represents the bounding box as four points defined in the image coordinate space using the DICOM SR SCOORD (Spatial Coordinates) content item. While the original JSON representation does not include any metadata to describe the nature of the finding being annotated, TID1500 relies on concept-value code pairs and SNOMED-CT (SCT)[15] terminology to unambiguously describe the semantics of the finding and its anatomic location (finding site). To encode annotations from the tool-specific JSON representation into TID1500, we developed a conversion script that uses the open-source `highdicom`[27] library. Given the references to the annotated DICOM images provided in the original JSON dataset, we rely on the DICOM metadata available in those images to extract and initialize composite context (patient- and study-level metadata) in the DICOM SR objects. We assign each bounding box “Lesion” (SCT code 52988006) as the finding, and “Lung” as the finding site (SCT code 39607008), while encoding the coordinates of the bounding box vertices exactly as they are available in JSON (we note that some entries in the JSON file were missing a bounding box definition and were skipped during conversion). As a result of the conversion, a single TID1500 instance (file) was created for each DICOM series annotated. In total, 970 TID1500 instances were generated, containing 9,280 bounding box annotations for 581 patients and 970 CT series.

## NLSTseg

The NLSTseg dataset[9] contains volumetric annotations (voxel segmentations) of the lung cancer lesions for patients who screened positive for lung cancer. The annotation protocol used the lung lobe(s) location of the lesion reported in the NLST clinical data as a guide for localizing the tumor region. According to the manuscript introducing the dataset, segmentations were performed by “a researcher with two years of imaging research experience and a radiologist specializing in oncology with five years of experience”, and were subsequently verified by “a radiation oncologist with ten years of experience and a radiologist with ten years of experience”[9]. The segmentations were originally saved in the NIfTI format, which is a volumetric image format popular in the research community but lacking metadata and deficient in interoperability beyond research workflows. In total, 662 tumors and 53 nodules from 605 patients were included in the dataset[32]. The original segmentations are accompanied by three Excel spreadsheets containing image-, patient-, and label-level metadata information. The image-level dataset details the referenced CT volume and acquisition parameters, such as instance IDs, convolution kernels, and manufacturer specifications. At the patient level, the data captures demographic profiles, including age and gender, alongside clinical diagnoses, such as the lung cancer subtype. The label-level data provides granular lesion characteristics, identifying whether a lesion is a tumor or a nodule, and specifying its lung lobe location.

We converted the latest version, v3, of the dataset, while discarding data for four patients due to data inconsistencies detailed below. Three `SeriesInstanceUIDs` were repeated for different PatientIDs and were therefore removed. One additional patient was removed from further analysis as the number of rows in the label spreadsheet did not match the number of unique labels in the lesion file, resulting in a total of 601 patients analyzed.

NIfTI segmentations were losslessly converted into DICOM SEG representation using the `dcmqi itkimage2segimage`[28] command-line converter. DICOM SEG composite context metadata was initialized from the referenced DICOM CT. The `AnatomicRegionSequence` DICOM attribute was used to encode the lung lobe location of the finding (details of the mapping are available in the companion source code; see Table 2), based on the metadata accompanying the NLSTseg segmentations. Designation of the finding as a tumor or nodule was captured in the `SegmentedPropertyTypeCodeSequence` attribute. Both of those relied on SNOMED-CT for codes. Multiple lesions segmented in the same CT series were encoded as separate segments within the same segmentation object. The resulting DICOM dataset contains 601 DICOM SEG series, corresponding to a total of 705 lesions segmented in 601 CT series for 601 patients.

To further increase the reusability and accessibility of the segmentations, we calculated quantitative features for the segmented regions. Specifically, we extracted first-order (i.e., summary statistics for signal intensity within the segmented region) and shape (i.e., various quantitative descriptors summarizing the geometry of the segmented region) features from each of the generated NLSTseg segmentations using the `pyradiomics`[33] package. DICOM TID1500 SRs were generated for each of these sets of features using the `highdicom` package[27] and the `tid1500writer` converter from the `dcmqi` package[28]. Image Biomarker Standardization Initiative (ISBI)[34] terminology was used for encoding the extracted feature quantities, while Unified Codes for Units of Measure (UCUM)[35] codes were used for describing the measurement units, per the established DICOM practice. For more information about the features extracted, please see "Radiomic feature mapping" in Table 2 of the Usage Notes. For one of the cases, the segmentation volume and referenced CT volume were of different dimensions, causing feature extraction from `pyradiomics`[33] to fail. Therefore, DICOM SR TID1500 objects were generated for a total of 600 patients.

### TotalSegmentator

To enrich the NLST collection with anatomic organ segmentations, we used TotalSegmentator[17], a deep learning model capable of segmenting over 100 anatomical regions in CT images, trained on a diverse dataset of patients with multiple pathologies, and collected from different scanners and institutions. As described in detail by Thiriveedhi et al.[36], the analysis was performed on 203,087 CT series from the NLST collection. We first applied a custom Google BigQuery SQL query to the DICOM metadata available in the BigQuery tables maintained by IDC to exclude NLST CT series that are not suitable for TotalSegmentator analysis (e.g., those that contain missing slices, have inconsistent resolution or orientation, including scout/localizer series), resulting in 126,088 CT series from 71,661 studies corresponding to 26,194 patients. Next, we used `dcm2niix`[29,30] to convert the CT images into an intermediate NIfTI representation as required by TotalSegmentator. We then segmented each of the converted CT series using TotalSegmentator v1.5.6, which produced segmentations in the NIfTI format, and extracted first-order and shape features using `pyradiomics`[33] for each segmented structure. The approach is described in detail in Thiriveedhi et al.[36].

TotalSegmentator[17] relies on numeric labels in the NIfTI files and their mapping to text strings for defining the semantics of the segmented structure. We mapped those tool-specific text labels to SNOMED-CT codes and used those for segment-specific metadata in DICOM SEG. Mapping of the TotalSegmentator labels to coded concepts is available under "Anatomical feature mapping" in Table 2 of the Usage Notes. Generation of both the DICOM SEG containing segmentation results and the TID1500 SR with the corresponding radiomics features followed the same approach and tools as the one described for the NLSTseg dataset. Due to conversion and geometry issues, 37 series were removed for processing. Therefore, in total, 126,051 DICOM SEG and 252,102 DICOM SR TID1500 instances were generated for 24,194 patients and 126,051 CT series[36].

# Data Records

The resulting datasets are provided as DICOM Segmentation and Structured Report objects to ensure the FAIR representation of the data and associated metadata. The data is organized into the following publicly available collections, which are accompanied by Zenodo data descriptors and can be accessed from the IDC at the respective URLs:

1. **NLST-Sybil**[37],
   https://portal.imaging.datacommons.cancer.gov/explore/filters/?analysis_results_id=NLST-Sybil
2. **NLSTSeg**[38],
   https://portal.imaging.datacommons.cancer.gov/explore/filters/?analysis_results_id=NLSTSeg
3. **TotalSegmentator-CT-Segmentations**[39],
   https://portal.imaging.datacommons.cancer.gov/explore/filters/?analysis_results_id=TotalSegmentator-CT-Segmentations

# Data Overview

NLST-Sybil contains one DICOM TID1500 SR object for each annotated CT series. NLSTSeg and TotalSegmentator-CT-Segmentations collections contain one DICOM SEG object for each CT series segmented. Each segmentation object is accompanied by a DICOM TID1500 SR object containing first-order radiomics features, and another SR containing shape radiomics features extracted from the segmentation. Metadata associating the segmentations, annotations, and radiomics features with the CT images and patients is communicated using standard DICOM attributes, as demonstrated in the Usage Notes. Please refer to Table 1 for further information.

| | NLST-Sybil | NLSTSeg | TotalSegmentator-CT-Segmentations |
|---|---|---|---|
| **# Patients** | 581 | 601 | 26,194 |
| **# DICOM studies** | 964 | 601 | 71,660 |
| **# DICOM CT series** | 970 | 601 | 126,051 |
| **# DICOM SEG series** | — | 601 | 126,051 |
| **# Shape radiomics features DICOM SR TID1500 series** | — | 600 | 126,051 |
| **# First-order radiomics features DICOM SR TID1500 series** | — | 600 | 126,051 |
| **# Planar annotations DICOM SR TID1500 series** | 970 | — | — |
| **Total # of lesions/bounding boxes** | 9,280 | 705 | — |

***Table 1***: *Summary of the content of each of the three datasets.*

# Technical Validation

## DICOM conformance

Conformance to the standard is critical to ensuring interoperability of DICOM content. We used established tools to evaluate standard compliance. We first validated the SEG and SR objects using the `dciodvfy` tool from `dicom3tools`[40], which verifies conformance of the object to the DICOM Information Object Definition (IOD). Further, we used the `DICOMSRValidator` command-line tool from `PixelMed Java DICOM Toolkit`[41] to establish conformance of the structured report content tree with the DICOM TID1500 SR template.

## Qualitative verification

To support visualization and aggregated access to the metadata across the collections in this data descriptor, we relied on several services available in Google Cloud. We first imported the DICOM content into a Google Healthcare (GHC) DICOM Store, which is a hosted service that implements efficient storage and querying of DICOM data via the DICOMweb interface[14]. We then used the GHC DICOM Store BigQuery export feature to make all of the DICOM metadata searchable using the Standard Query Language (SQL) interface. The resulting BigQuery table makes it straightforward to use queries to ensure the completeness of data, the veracity of various metadata attributes, and to enable data exploration.

To establish a visual interface for navigating the data, we developed several Google Data Studio dashboards[42] offering various charts and controls to subset and scrutinize the data. Specifically, the dashboards include summaries and controls for key attributes of the data, such as the annotated finding, image acquisition, and clinical characteristics. The dashboards include URLs for the individual DICOM studies that can be used to visualize segmentations and planar annotations in the IDC-hosted OHIF Viewer[43], which retrieves the DICOM data from the IDC-maintained DICOM store mentioned earlier. Our overall strategy for verification was to examine the distribution of attributes for the segmentations/annotations (e.g., the volume of the segmented region, lung lobe location of a tumor, etc), and selectively review visually outlier cases as defined by that distribution. Further, images and annotations for the individual cases can be examined using the 3D Slicer[44], with the QuantitativeReporting extension enabling loading of the DICOM SEG/SR objects via the `dcmqi` library[28]. We next briefly discuss dataset-specific examination and verification of the data. Please refer to Figure 1 to see a flowchart for the workflow and Figure 2 for various examples of the annotations generated.

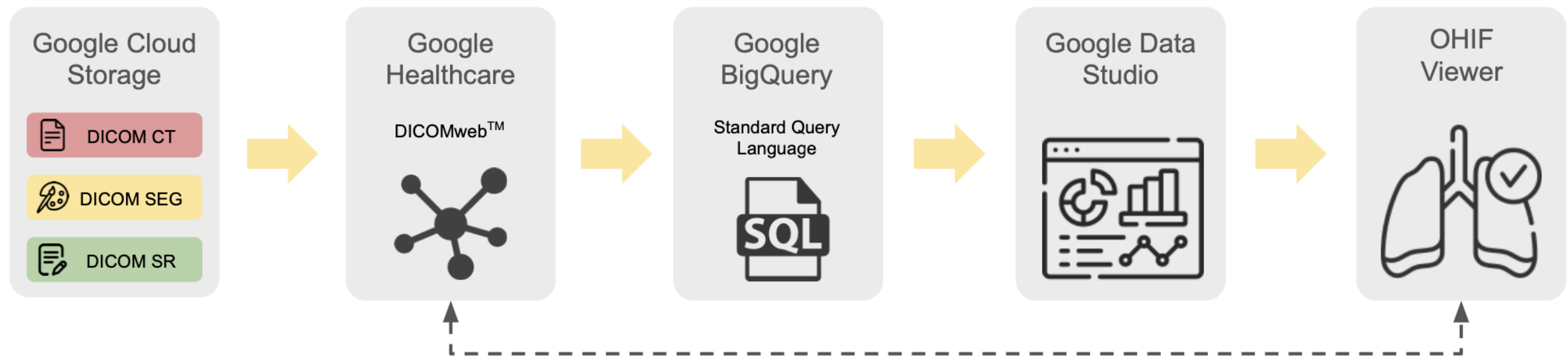

***Figure 1: Flowchart for exploring and visualizing the data and annotations.** DICOM files from Google Cloud Storage are first ingested into Google Healthcare. Metadata is then extracted using Google BigQuery into tables, which power interactive dashboards using Data Studio. The OHIF viewer is integrated to allow for visualization of the data and annotations.*

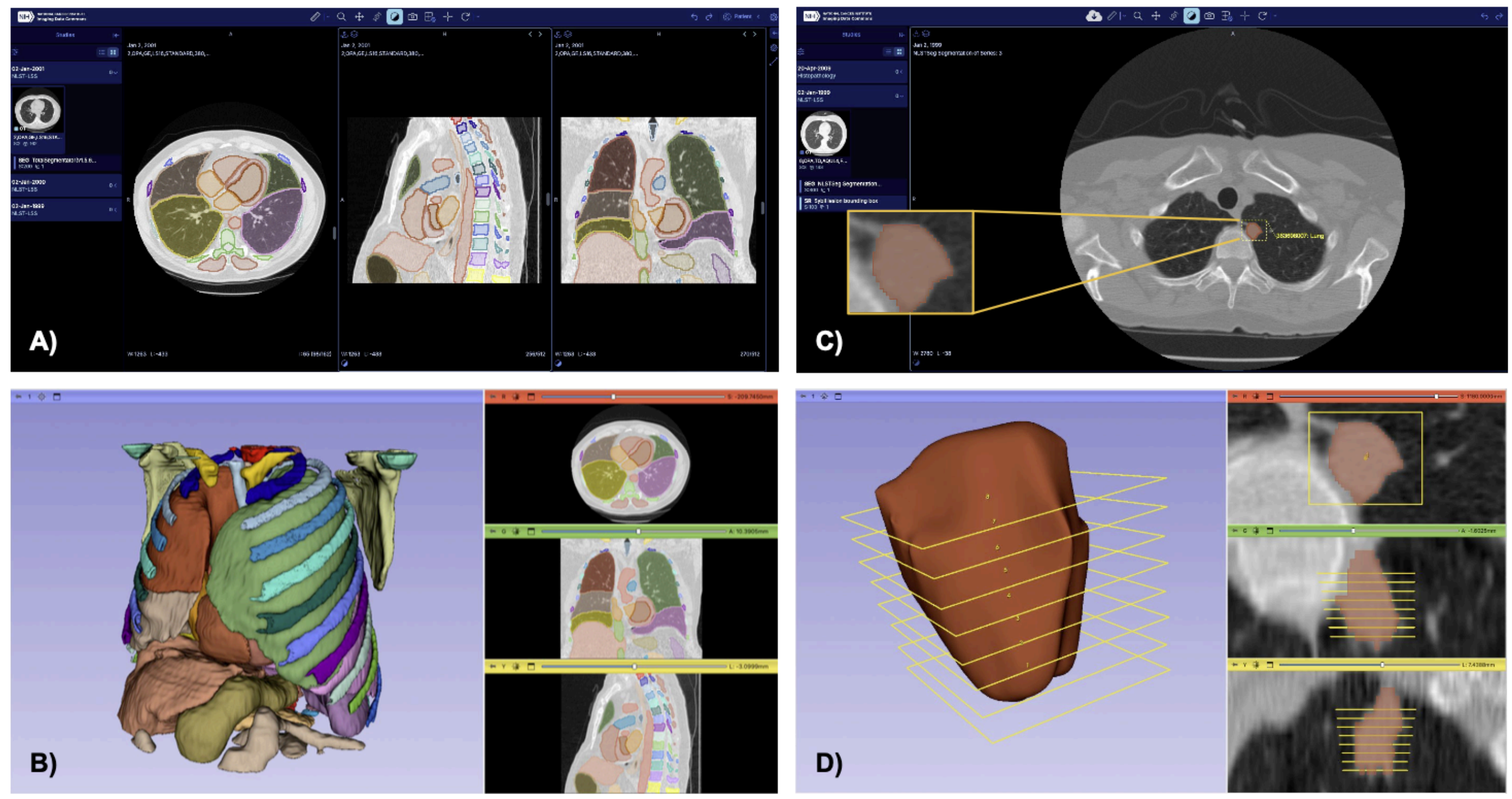

***Figure 2: Examples of visualizations that were used to support qualitative verification of the data.** A) TotalSegmentator segmentations in OHIF, B) TotalSegmentator segmentations in 3D Slicer, C) NLSTSeg lesion segmentation and NLST-Sybil bounding box in OHIF, and D) NLSTSeg lesion segmentation and NLST-Sybil bounding boxes in 3D Slicer.*

**TotalSegmentator-CT-Segmentations** The dashboard for this dataset (page 1 of the dashboard link in Table 2) leverages the quantitative features derived from the segmented structures. As the dataset contains thousands of anatomical regions with no expert ground truth available, being able to select and filter out patients based on criteria in the dashboard is crucial for the identification of outliers and other problematic cases. To visually verify the segmentations, we sampled extreme points from the distributions of volume values and examined the corresponding cases visually in OHIF. The example shown in Figure 3 demonstrates filtering for the volume of the middle lobe of the right lung and selecting a patient with a low volume. Upon further inspection, the low volume is due to incomplete coverage of the anatomy by the CT scan. Additional strategies for evaluating segmentations in the absence of ground truth are discussed in Krishnaswamy et al[45].

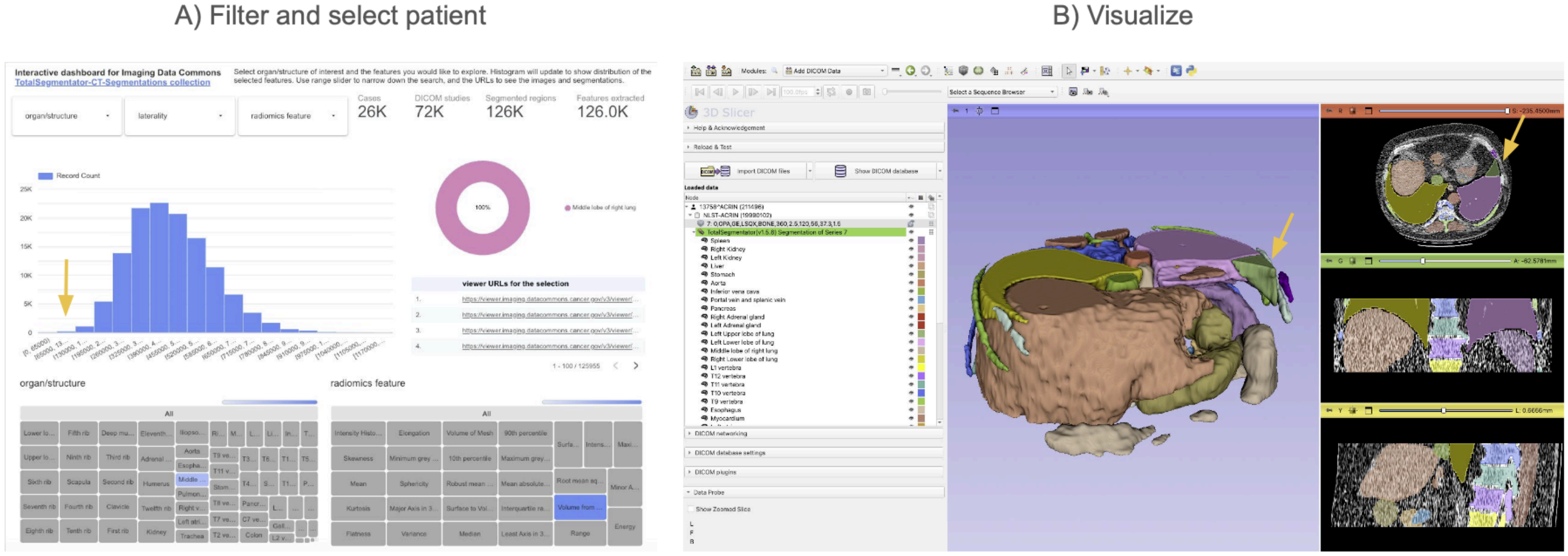

*__Figure 3__: __Example of examining outlier patients.__ A) First, we filter for a particular organ and select the middle lobe of the right lung. Then, we choose a particular radiomics feature: the voxel from the volume summation. We then select a patient with a low volume, as indicated by the yellow arrow. B) Next, we visualize the patient in 3D Slicer and verify that the outlier is due to a cropped CT scan.*

**NLST-Sybil** The dashboard prepared for this dataset (page 2 of the dashboard link in Table 2) focuses on the assessment of the characteristics of the tumor bounding box annotations. The dashboard enables examination of the distribution of the number of bounding boxes per group, along with the areas of the individual annotations, and identification of those CT series that have multiple bounding boxes per slice. These components of the dashboard help quickly locate cases that could be potentially problematic for visual verification.

The dashboard links tumor characteristics available from NLST clinical data (lung lobe location of the tumor, margins, predominant attenuation) and patient characteristics (lung cancer type and lung cancer stage) with the individual annotations (see page 3 of the dashboard link in Table 2). This clinical metadata available from NLST is included for the axial slice where the lesion has the largest diameter. Please refer to the SQL queries in the Code Availability for further details on how the linkage was established. Figure 4 demonstrates an example of correlating the location of the segmented tumor with the information reported in the NLST clinical metadata. We first filter for tumors in the right upper lobe, and after selection, we visualize the bounding boxes in OHIF. Using the multi-planar reconstruction (MPR) visualization mode, we can confirm that the tumor is located in the right upper lobe.

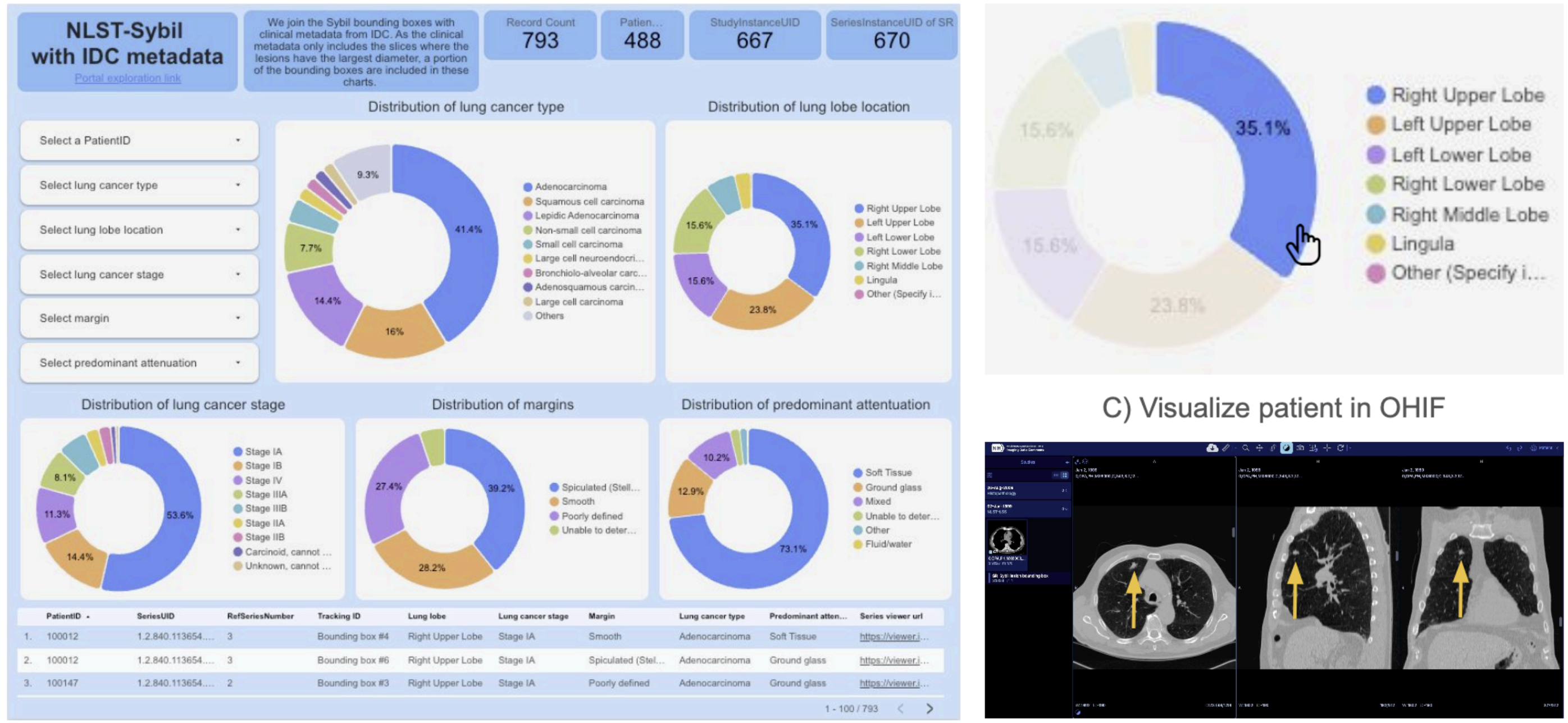

***Figure 4**: Examples of technical verification of the Sybil dataset joined with clinical metadata. A) demonstrates the distributions of the various metadata, where we can then B) filter for specific patients by selecting a lobe (right upper lobe). After filtering, we can C) visualize the selected patient in OHIF and verify that the tumor is indeed present in the right upper lobe.*

**NLSTSeg** The dashboard prepared for this dataset (pages 4-6 of the dashboard link in Table 2) focuses on the assessment of the characteristics of the lesion segmentations. This dashboard enables examination of the segmentations and their associated metadata, for instance, analyzing tumors vs nodules and examining the lung lobe locations of the lesions. A histogram summarizing segmentation-derived features allows the user to quickly identify potentially problematic lesions. The dashboard links lesions with the accompanying clinical data, including lung cancer staging, histology, and tumor margin characteristics (see page 6 of the dashboard link in Table 2). Figure 5 demonstrates an example of verification of the tumor margins (spiculated, poorly defined, smooth, etc) vs the lung cancer stage, allowing the user to examine the actual margins of the segmented tumor and compare with the characteristics reported in the clinical data.

We utilized `pyradiomics`[33] to extract first-order and shape features from the lesion segmentations (page 5 of the dashboard link in Table 2). In the original dataset, the authors included the volume of the lesion in the provided Excel files. To validate the reported values, we used `pyradiomics`[33] library to re-calculate tumor volume and confirmed those agree with the earlier reported values (Pearson correlation coefficient of 0.99, mean difference of $1.92*10^{-5} \pm 0.0029$ mL). See Colab notebook `validateNLSTSegVolume.ipynb` in the GitHub repository for further details.

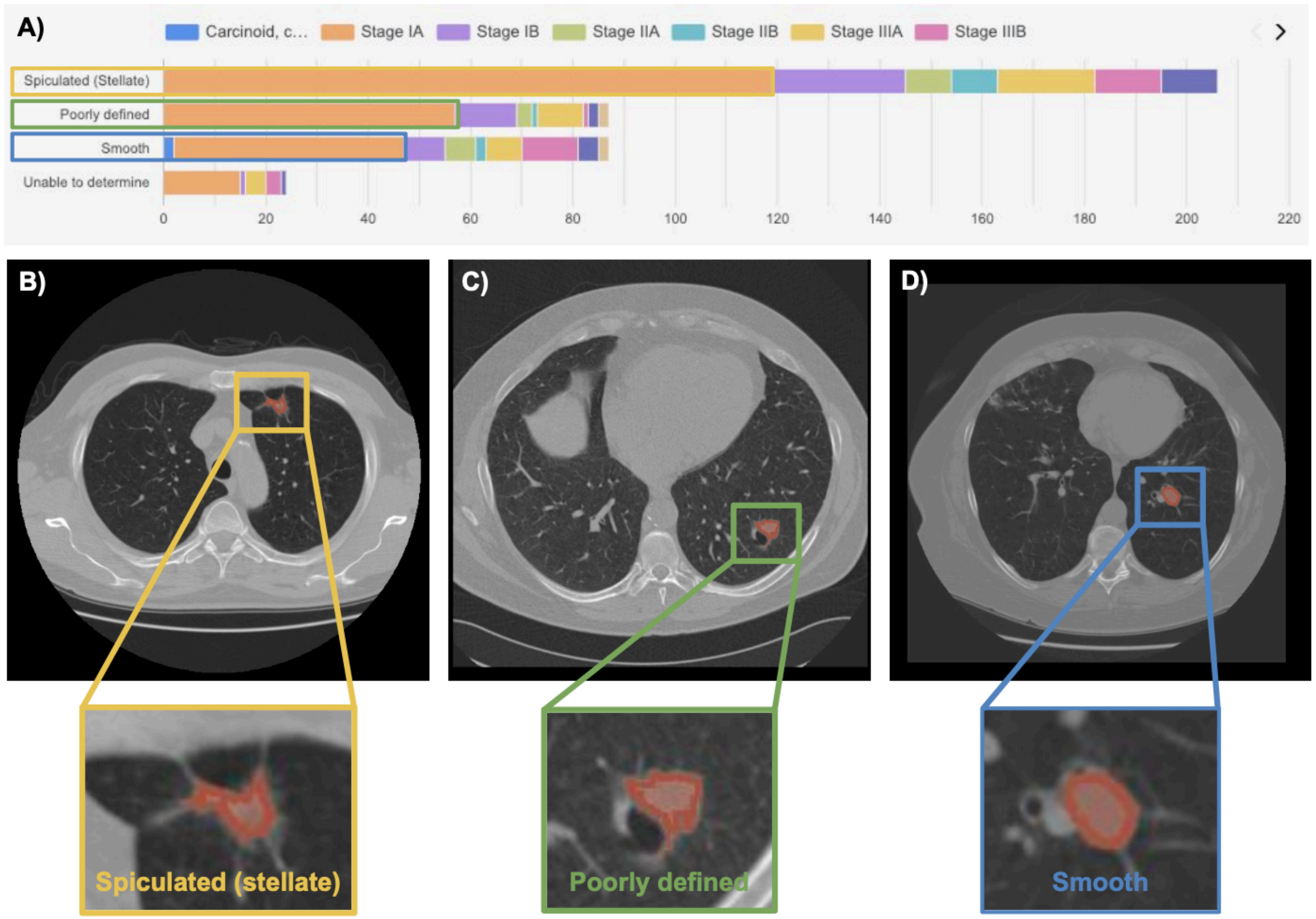

***Figure 5**: **Examples of verification of the tumor margin for three patients using the dashboard.** A) The top row shows the distribution of patients according to stage, and B), C), and D) show three examples of different tumor margins. We can visually see that the tumor with a spiculated margin indeed has irregular and spiky edges, and that the tumor with a smooth margin has smooth and well-defined edges.*

# Cross-collection consistency evaluation

Availability of complementary annotations in a uniform representation makes it possible to combine them to gain insight into both their quality and the images they annotate. To demonstrate this capability and to support future analyses of the data, we developed several dashboards that span multiple annotation collections. First, we calculated the overlap between the volumetric segmentations in NLSTSeg and bounding box annotations in NLST-Sybil and developed a dashboard to explore the agreement between these two independently generated annotations (page 7 of the dashboard link in Table 2). Second, we developed a dashboard that utilizes TotalSegmentator's automatic lung lobe segmentations to evaluate the assignment of the lesions available in the NLST clinical data (page 8 of the dashboard link in Table 2). Being able to quickly flag lesions where such an assignment is inconsistent can help with quality control of the annotations and potentially can help identify images where AI segmentation fails.

**Sybil vs NLSTseg agreement** We compare the lesion bounding boxes from Sybil and the lesion segmentations from NLSTSeg (page 7 of the dashboard link in Table 2). For the series that have overlapping bounding boxes and segmentations, the goal was to verify that the NLSTseg lesion was within the Sybil bounding box. We first filtered for series that included both a bounding box and a lesion segmentation. Next, we identified the individual slices that also contained both a bounding box and lesion segmentation. Then, we computed the overall fraction of tumor voxels inside all boxes for a particular series by summing over the individual slices (see the Google Colab notebook `NLSTSegVsNLSTSybil.ipynb` in the GitHub repository for further details). To assess out-of-plane annotation discrepancies, we evaluated precision and recall across slices. Precision was defined as the proportion of bounding box slices that also contained a segmentation, while recall was defined as the proportion of segmentation slices that contained a bounding box (see `NLSTSegVsNLSTSybil_viz.ipynb`). Because precision was consistently 1.0 across this set of patients, indicating that every slice with a bounding box included a corresponding segmentation, we focused our analysis on recall. See Figure 6 for distributions of the tumor fraction and recall, and two examples comparing the segmented tumor volume from NLSTSeg with the Sybil bounding box annotations.

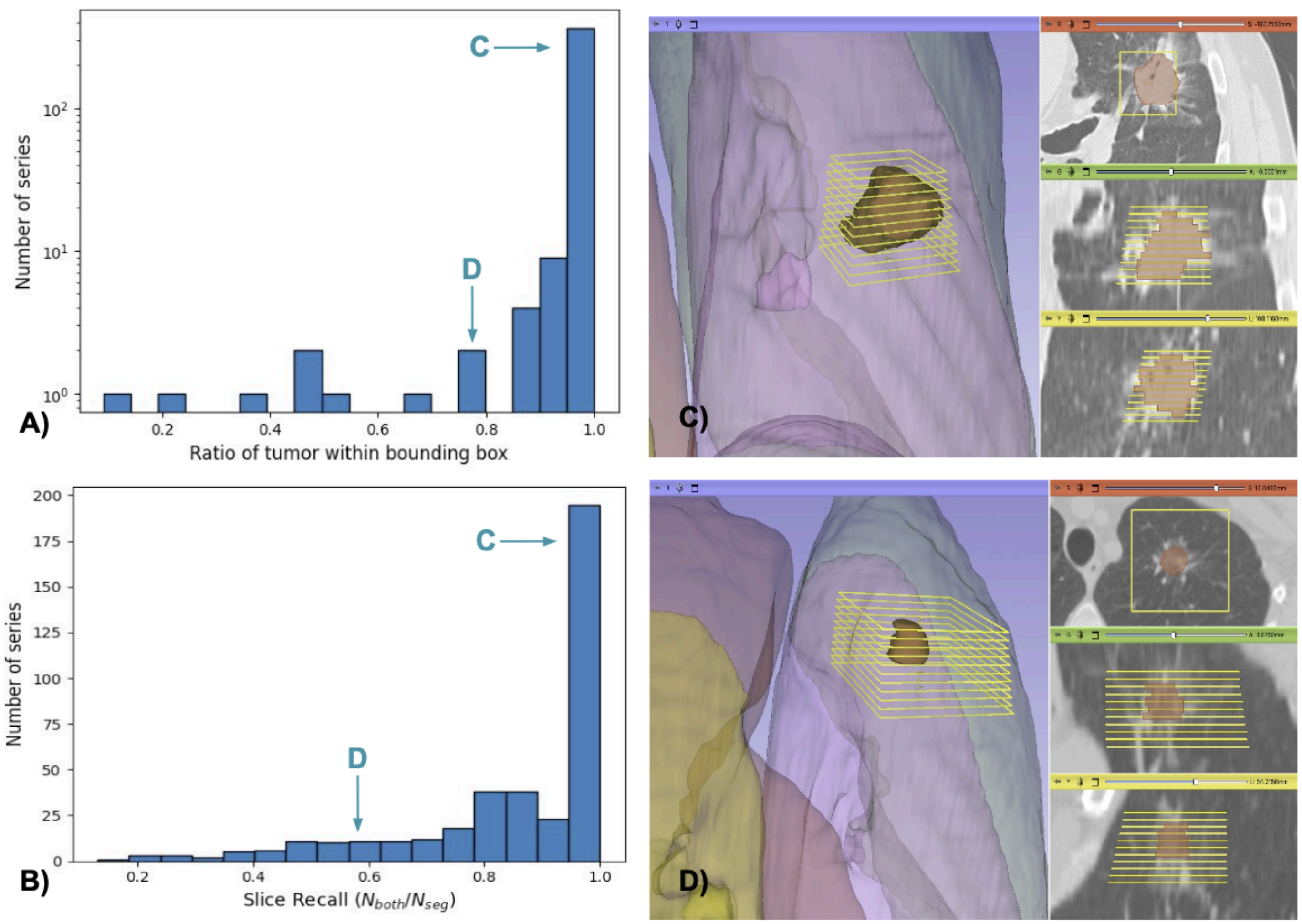

***Figure 6: Examples of comparisons of tumor segmentations in NLSTSeg and bounding boxes in Sybil, with TotalSegmentator-generated segmentations in 3D Slicer.*** *A) Tumor fraction analysis performance evaluating the fraction of the tumor that overlaps with the bounding box area. B) Recall metric evaluating cross-sectional, out-of-plane segmentation completeness relative to bounding box extents across axial slices. A recall approaching 1.0 indicates tight axial slice alignment between automated segmentations and bounding box limits, whereas lower recall values demonstrate out-of-plane boundary discrepancies where bounding box annotations exist without corresponding segmented voxels. C) High-concordance example (tumor fraction = 0.95, recall = 1.00) demonstrating complete axial slice-wise overlay where the segmented lesion occupies a substantial volumetric fraction of the bounding box volume without missing inter-slice masks. D) Low-concordance example (tumor fraction 0.79, recall = 0.583) exhibiting a slice-wise mismatch, where the Sybil bounding box spans additional axial slices that lack corresponding NLSTSeg*

*masks. This reflects either strict segmentation boundaries or expanded bounding box margins.*

**NLSTSeg vs TotalSegmentator-CT-Segmentations lung location agreement** TotalSegmentator-CT-Segmentation includes volumetric segmentation of the individual lung lobes, allowing for independent verification of the location of the lesion provided by NLSTSeg (see page 8 of the dashboard link in Table 2). For the CT series that were annotated by both collections (n=575), the user can filter lesions that are assigned a lung lobe location that does not agree with the TotalSegmentator segmentation results, as illustrated in Figure 7. Approximately 26% of the tumor locations from NLSTSeg do not agree with the TotalSegmentator-reported location, as seen in Figure 8 (see the Google Colab notebook `NLSTSegVsTS_viz.ipynb`).

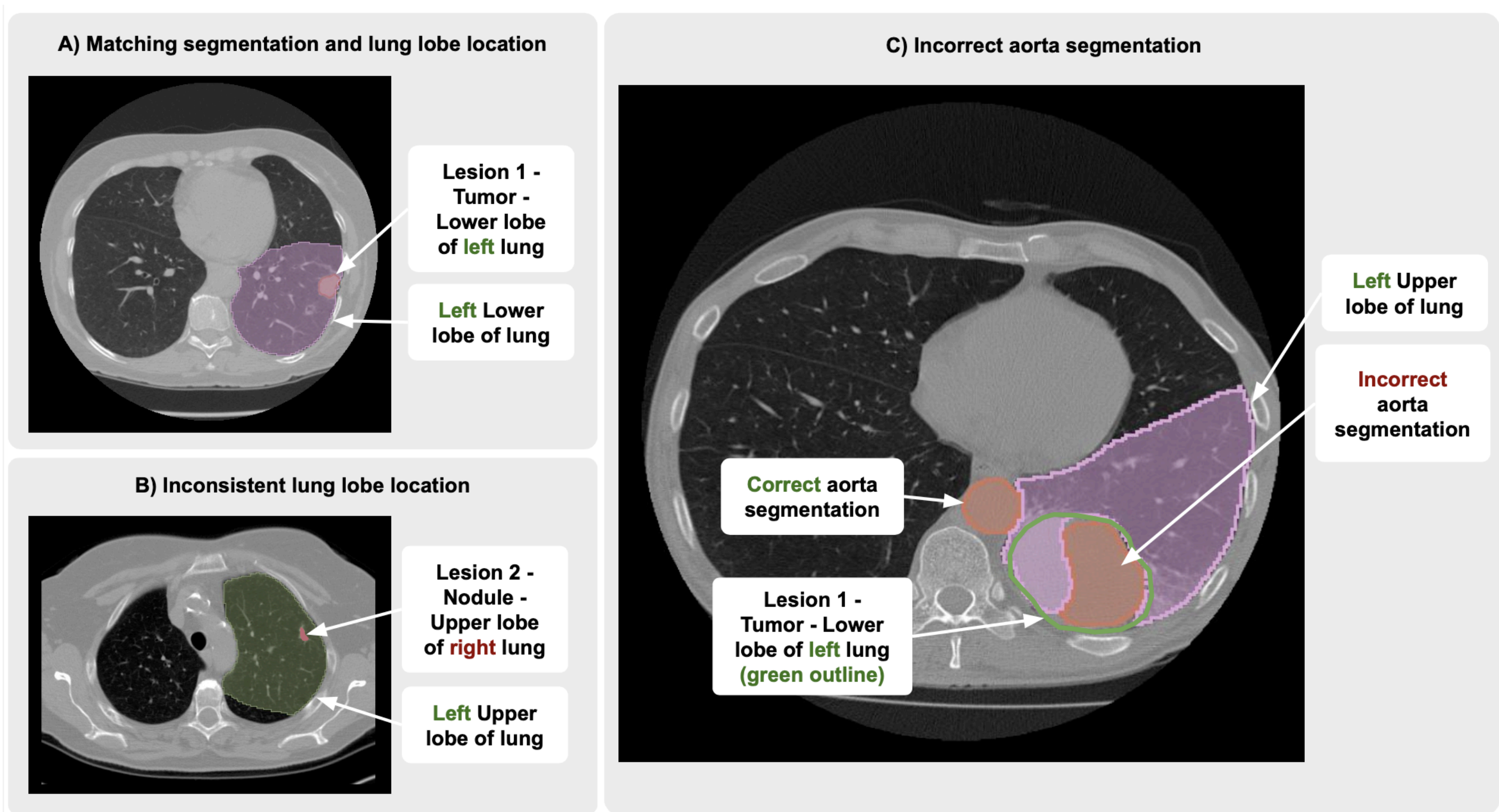

***Figure 7: Examples of comparisons of tumor location reported in NLSTSeg and deduced from lung lobe segmentations produced by TotalSegmentator.*** *Red text color highlights errors. A) Example of a tumor where the lung lobe label matches the TotalSegmentator segmentation result, B) Example of a nodule with inconsistent lung lobe location, and C) Example of a patient where the aorta segmentation from TotalSegmentator incorrectly includes a portion of the tumor.*

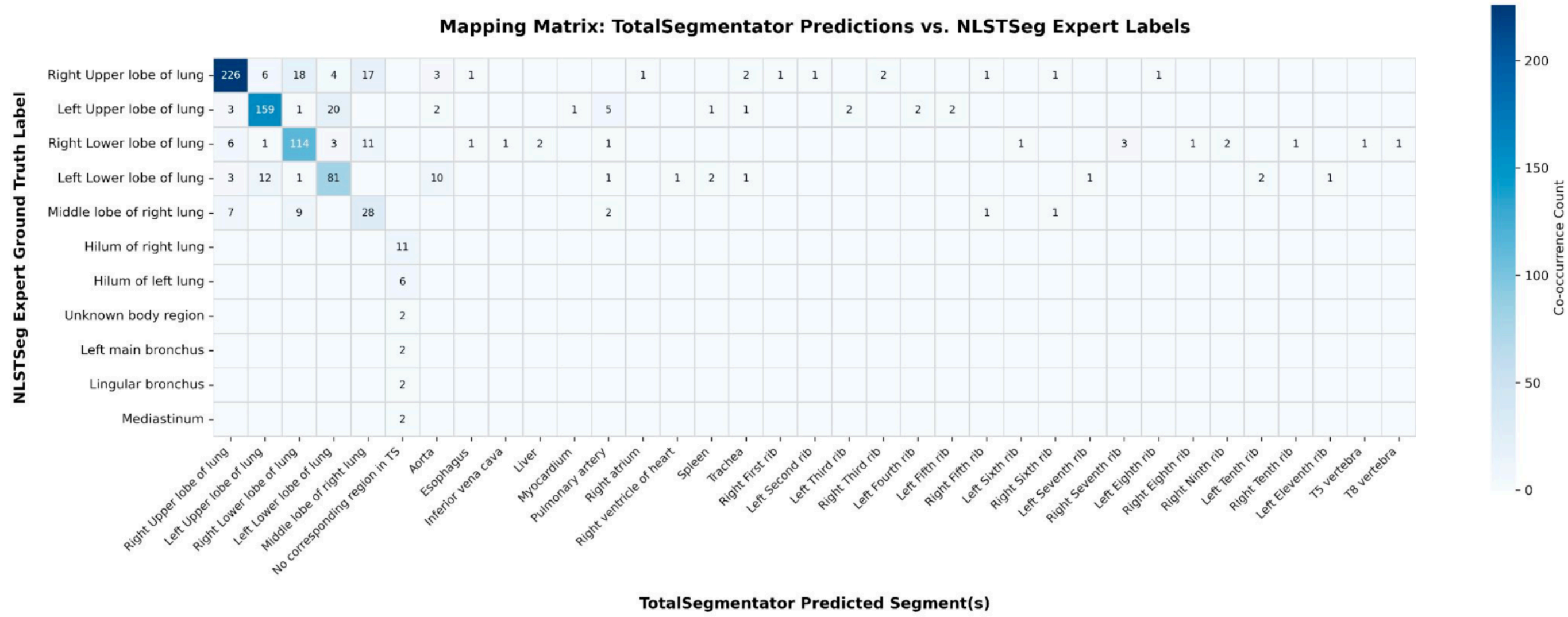

***Figure 8: Comparison of the tumor location based on the segment assignment by TotalSegmentator vs the NLSTSeg expert-assigned anatomic location.*** *Numbers indicate the total number of segments, where the darker colors indicate a higher number of segments. Note the high overlap in corresponding lung regions; however, there are many extraneous lung regions, cardiac substructures, and ribs segmented by TotalSegmentator that overlap with the segmented lesion and disagree with the NLSTSeg expert assignment of lesion location.*

# Usage Notes

In this section, we demonstrate how to visualize and use the generated standardized annotation datasets, including 1) TotalSegmentator segmentations and segmentation-derived features, 2) Sybil bounding boxes, and 3) NLSTSeg lesion segmentations and segmentation-derived features:

- Visualization in Slicer: The `DICOMTID1500 plugin` was modified to load and view multiple graphic types and planar annotations from the SR files, including 2D bounding boxes, 3D points, and 2D lines. Each of these types of annotations was loaded as a markup: `vtkMRMLMarkupsROINode`, `vtkMRMLMarkupsFiducialNode, and vtkMRMLMarkupsLineNode`, respectively.

- Visualization in OHIF: OHIF[43] (https://github.com/ImagingDataCommons/ViewersV3) was utilized to view and interact with the generated DICOM files. Functionality was added in order to read and display the planar annotations in the SR format.

- Parsing the DICOM objects: We provide a Google Colab notebook that demonstrates how to read, parse, and display DICOM Segmentation and Structured Report objects: https://github.com/ImagingDataCommons/idc-nlst-plus/blob/main/UsageNotes/parseSEGandSR.ipynb. Using `dcmqi`[28] and `highdicom`[27], we demonstrate how to parse the DICOM files and obtain the referenced images for enhanced visualization of the lesion.

We additionally provide links to various resources in Table 2.

| Name | Usage | URL |
|---|---|---|
| NLSTSeg codes | Maps the lesion type to coded concepts | https://github.com/ImagingDataCommons/idc-nlst-plus/blob/main/NLSTSeg_codes.csv |

| Radiomic feature mapping | Maps the radiomics features to coded concepts | https://github.com/ImagingDataCommons/CloudSegmentator/blob/main/workflows/TotalSegmentator/resources/radiomicsFeaturesMaps.csv |
|---|---|---|
| Anatomical feature mapping | Maps the anatomical regions to coded concepts | https://github.com/ImagingDataCommons/CloudSegmentator/blob/main/workflows/TotalSegmentator/resources/totalsegmentator_snomed_mapping_with_partial_colors.csv |
| Dashboard | Interactive dashboards to filter, search, and visualize the data | https://lookerstudio.google.com/s/qWzfyzQJQsY |

***Table 2**: Resources including look-up tables and dashboards.*

# Dataset-specific usage notes

## TotalSegmentator-CT-Segmentations

We performed a study investigating the use of heuristics to quickly filter segmentations for cohort building[45]. In this study, we developed multiple heuristics to measure the consistency of the segmentations, including the presence of complete organ segmentations, verification of the laterality of the organ, and filtering based on the volume of the anatomical regions. Exploratory notebooks are available at https://github.com/ImagingDataCommons/CloudSegmentatorResults. A Streamlit application allows users to apply the heuristics to individual segmented structures. See here: https://huggingface.co/spaces/ImagingDataCommons/CloudSegmentatorResults.

## NLST-Sybil

To facilitate exploratory analysis and discovery within the NLST-Sybil cohort, we developed an interactive website to demonstrate content-based image retrieval based on the segmented tumor regions. We used embeddings generated by applying nine different foundation models to the tumor/nodule bounding box annotations[46] to construct a connectome - an interactive graph connecting nodes corresponding to the individual segmentations based on the cosine similarity between these high-dimensional representations. This interface integrates the embeddings with patient characteristics and clinical metadata, enabling researchers to explore data relationships and identify interesting clustering patterns. We make our code and interactive visualization publicly available at: https://github.com/ImagingDataCommons/nlst-sybil-connectome. Please refer to Figure 9 for an example of this visualization.

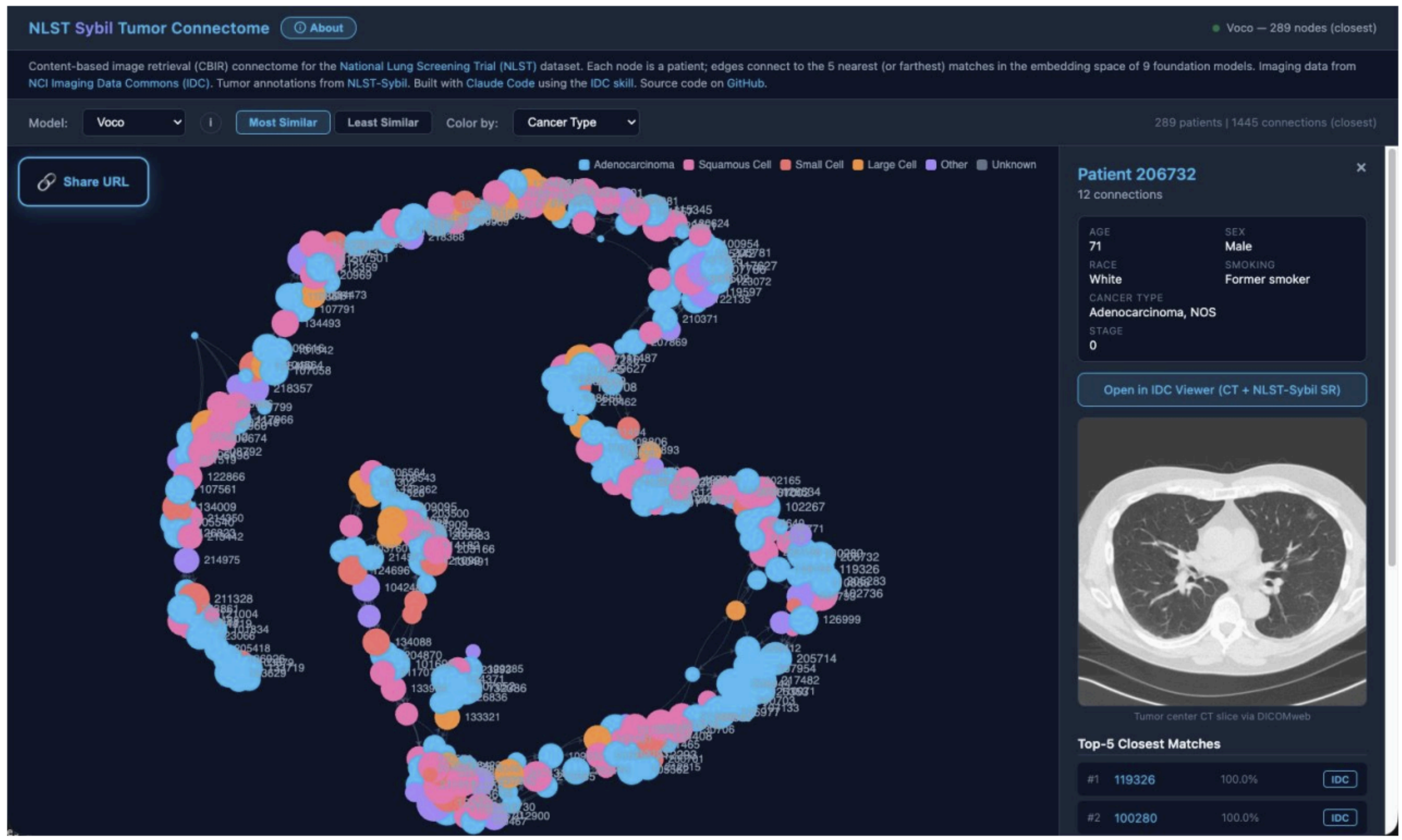

***Figure 9: Connectome visualization for segmentation-enabled content-based image retrieval.*** *Users can choose a specific foundation model and explore the pre-computed tumor embedding space using cosine similarity. Nodes (patients) are colored by patient or clinical characteristics and can aid the user in identifying various patterns.*

# Data Availability

The converted and generated datasets are available through Zenodo records and through the IDC portal, at the following URLs:

- TotalSegmentator-CT-Segmentations[39]:
  - https://zenodo.org/records/13900142
  - https://portal.imaging.datacommons.cancer.gov/explore/filters/?analysis_results_id=TotalSegmentator-CT-Segmentations
- NLST-Sybil[37]:
  - https://zenodo.org/records/15643335
  - https://portal.imaging.datacommons.cancer.gov/explore/filters/?analysis_results_id=NLST-Sybil
- NLSTSeg[38]:
  - https://zenodo.org/records/17362625
  - https://portal.imaging.datacommons.cancer.gov/explore/filters/?analysis_results_id=NLSTSeg

# Code Availability

Our code is available in the following GitHub repository: https://github.com/ImagingDataCommons/idc-nlst-plus, where the code is available under a permissive MIT license. The code is organized as follows:

```
idc-nlst-plus/
├── DataRecords
│   ├── createNLSTSeg.ipynb
│   └── createNLSTSybil.ipynb
├── LICENSE
├── NLSTSeg_codes.csv
├── NLSTSeg_lung_codes.csv
├── README.md
├── SQL
│   ├── page1-totalsegmentator
│   │   └── quant_seg_viewer_view.sql
│   ├── page2-nlst-sybil
│   │   ├── nlst_sybil_bbox_measurements.sql
│   │   ├── nlst_sybil_bbox_measurements_with_urls_and_counts_view.sql
│   │   └── nlst_sybil_measurement_groups.sql
│   ├── page3-nlst-sybil-with-idc-metadata
│   │   └── nlst_sybil_bbox_measurements_with_idc_data.sql
│   ├── page4-nlstseg-explore-segmentations
│   │   └── nlstseg_segmentations.sql
│   ├── page5-nlstseg-features
│   │   ├── nlst_quantitative_measurements.sql
│   │   └── nlstseg_quantitative_measurements_with_minmax_volume.sql
│   ├── page6-nlstseg-with-idc-metadata
│   │   └── nlstseg_segmentations_and_meas_with_idc_data.sql
│   ├── page7-nlstseg-vs-nlst-sybil
│   │   └── info.txt
│   └── page8-nlstseg-vs-totalsegmentator
│       ├── info.txt
│       └── nlstseg_ts_lesion_matching_with_volume.sql
├── TechnicalValidation
│   ├── consistencyChecks
│   │   ├── NLSTSegVsTS.ipynb
│   │   ├── NLSTSegVsTS_viz.ipynb
│   │   ├── nlstseg_ts_lesion_matching_with_volume.csv
│   │   ├── nlst_sybil_and_nlstseg_overlap_per_series.csv
│   │   ├── NLSTSegVsNLSTSybil.ipynb
│   │   └── NLSTSegVsNLSTSybil_viz.ipynb
│   ├── technicalCompliance.ipynb
│   └── validateNLSTSegVolume.ipynb
├── UsageNotes
│   └── parseSEGandSR.ipynb
└── requirements.txt
```

# Author Contributions

Deepa Krishnaswamy was responsible for the generation of DICOM Segmentation and Structured Report objects, the creation of the dashboards, the development of the Google Colab notebooks, and the creation of the GitHub repository. Deepa was the main contributor in terms of writing, preparing text and figures, and editing this manuscript. Vamsi Thiriveedhi was responsible for the creation of the DICOM objects for TotalSegmentator, the queries, and notebooks. Suraj Pai was responsible for the development of the Tumor Imaging Benchmark, used to extract foundation model embeddings from the bounding box annotations (Sybil). David Clunie was responsible for the correct encoding of the DICOM Segmentation objects and Structured Reports. Igor Octaviano was responsible for the OHIF viewer: parsing and the display of the DICOM SR planar annotations. Christopher P. Bridge assisted with the creation of the DICOM objects using the highdicom package. Steve Pieper was responsible for the formulation and editing of the manuscript. Ron Kikinis was responsible for the formulation and editing of the manuscript. Andrey Fedorov was responsible for the overall supervision of the project, formulation, and structure of the manuscript, review of code, and extensive editing.

# Competing Interests

The author(s) declare no competing interests.

# Funding

This work has been funded in whole or in part with Federal funds from the National Cancer Institute, National Institutes of Health, under Task Order No. HHSN26110071 under Contract No. HHSN261201500003I.